\documentclass[english,aps,showpacs,superscriptaddress,floatfix, notitlepage,reprint,unicode=true,colorlinks=true,citecolor=Blue,linkcolor=RubineRed,urlcolor=Blue]{revtex4-1}
\usepackage[T1]{fontenc}
\usepackage[utf8]{inputenc}
\usepackage{babel}
\usepackage{mathrsfs}
\usepackage{bm}
\usepackage{amsmath}
\usepackage{amsthm}
\usepackage{amssymb}
\usepackage{graphicx}
\usepackage[]
 {hyperref}

\makeatletter
\theoremstyle{plain}
\newtheorem{thm}{\protect\theoremname}

\usepackage{braket}
\usepackage{bm}
\usepackage{graphicx}
\usepackage[usenames,dvipsnames]{xcolor}
\theoremstyle{definition}
\newtheorem{theorem}{Theorem}
\newtheorem{obs}[theorem]{Observation}
\usepackage{txfonts}
\usepackage{lettrine}
\date{\today}

\makeatother

\providecommand{\theoremname}{Theorem}

\begin{document}
\title{Lossless Postselected Quantum Metrology with Quasi-pure Mixed States}
\author{Jing Yang~\href{https://orcid.org/0000-0002-3588-0832}{\includegraphics[scale=0.05]{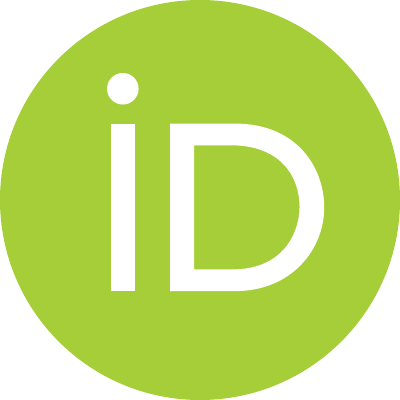}}}
\email{jing.yang.quantum@zju.edu.cn}

\address{Institute for Fundamental and Transdisciplinary Research, Institute
for Quantum Sensing, and Institute for Advanced Study in Physics,
Zhejiang University, Hangzhou 310027, China}
\address{Nordita, KTH Royal Institute of Technology and Stockholm University,
Hannes Alfv\'ens vag 12, 106 91 Stockholm, Sweden}
\begin{abstract}
Postselection can compress the metrological information and improve
sensitivity in the presence of certain types of technical noise. Postselected
quantum metrology with pure states has been significantly advanced
recently. However, extending this framework to mixed states leads
to formidable challenges, such as the difficulty in searching for
lossless postselection measurements or even the loss of metrological
information. In this work, we leverage the intuition for the lossless
postselection of pure states and generalize the theory to the lossless
postselection of a class of mixed states, dubbed quasi-pure states.
We illustrate our findings in postselected quantum imaging, unitary
estimation problems, and show that the quasi-pure structure can be
universally engineered through only classical correlation with an
ancilla. Our findings extend the utility of postselection techniques
to scenarios with decoherence and also offer new perspectives to foundational
questions in quantum information geometry.

\medskip{}
\end{abstract}
\maketitle
\lettrine[lraise=0.1, nindent=0em, slope=-.5em]{Q}uantum metrology
exploits the quantum coherence and quantum entanglement offered by
quantum mechanics for precision measurements and therefore promises
very high sensitivity. Pioneered by Helstrom~\citep{helstrom1976quantum},
Holevo~\citep{holevo2011probabilistic}, and many others~\citep{hayashi2005asymptotic},
quantum metrology has witnessed very rapid development thanks to the
advancement of quantum technology. In practice, a quantum sensor can
lose the quantum advantage beyond the coherence time due to its interaction
with ambient environments. Numerous efforts have been dedicated to
the study of the precision limits~\citep{escher2011general,escher2012quantum,chin2012quantum,demkowicz-dobrzanski2012theelusive,alipour2014quantum,demkowicz-dobrzanski2014usingentanglement,alipour2015extended,smirne2016ultimate,beau2017nonlinear,demkowicz-dobrzanski2017adaptive,haase2018precision,liu2017quantum,yang2019memoryeffects,altherr2021quantum,bai2023floquet,yang2024quantum}
for quantum sensing in noisy environments. 

In a parallel line, motivated by the quest for investigating anomalous
values of observables in quantum mechanics, Aharonov, Albert, and
Vaidman~\citep{aharonov1988howthe} propose to measure an observable
in a subensemble through pre- and post-selection, resulting in the
discovery of weak values of observables. The weak value of an observable
can lie far more outside the range of the spectrum of the observable
and therefore can be utilized to amplify weak signals, as experimentally
demonstrated in a variety of experiments~\citep{hosten2008observation,dixon2009ultrasensitive,starling2009optimizing,starling2010precision,xu2013phaseestimation}.
Recently, weak value metrology~\citep{dressel2014colloquium}, which
only concerns projective post-selection measurements, has been further
extended to generic post-selected quantum metrology, where general
measurements characterizes positive operator-valued measures (POVMs)
are considered~\citep{arvidsson-shukur2020quantum,lupu-gladstein2022negative,jenne2022unbounded,yang2023theoryof,yang2024quantum}.
Moreover, post-selected quantum metrology and standard quantum metrology
can be recast into a unified framework~\citep{yang2023theoryof}.
The key idea is that optimal measurements in standard quantum metrology
maximally extract the information about the estimation parameter into
the measurement statistics whereas those in post-selected quantum
metrology losslessly compress the complete information into a subensemble
with a small number of samples. This observation holds if the samples
are in pure states and it is not clear whether postselection of mixed
states can be made lossless or not. The primary technical challenge
in this problem is that the expression of the quantum Fisher information
(QFI) for the post-measurement states becomes formidable to evaluate
for mixed states. For example, Ref.~\citep{salvati2023compression}
concluded that if the depolarization channel is applied before the
post-selection measurement discussed in Refs.~\citep{jenne2022unbounded,lupu-gladstein2022negative},
then loss of the precision can occur. 

In this work, we adopt a physics and intuitive approach to the problem.
We analyze a binary-outcome lossless post-selection measurement on
a pure state $\rho(x)=\ket{\psi(x)}\bra{\psi(x)}$ where $x$ is the
estimation parameter. We observe that the effect of the post-selection
is to amplify the parametric derivative $\partial_{x}\rho(x)$ by
a large factor $1/\sqrt{\lambda_{\checkmark}}$, where $\lambda_{\checkmark}$
is the post-selection success probability, while the post-selected
state remains to be $\rho(x)$. Inspired by this physical observation,
we generalize the lossless post-selection strategy to a class of mixed
state $\rho(x)$, which is ``quasi-pure'' in the sense that $\partial_{x}\rho(x)$
behaves in a similar manner with pure states. Such quasi-pure states
possess elegant mathematical structures and therefore offers compact
evaluation of many quantities in quantum. Furthermore, our findings
also straightforwardly generalize to lossless postselection with multiple
estimation parameters and multiple outcomes. We then show that the
quasi-pure structure finds applications in postselected quantum imaging
and unitary estimation and can be engineered universally by creating
classical correlations of the probe system with an ancillary system.
Our theory provides a framework to apply postselection techniques
to suppress technical noise in the presence of decoherence and can
be applied to practical metrological tasks, such as realistic optical
imaging and distributed quantum sensing. At the fundamental level,
our findings also offer new perspectives to quantum information geometry. 

\medskip{}

\noindent{\large\textbf{Results}}{\large\par}

\begin{figure}
\begin{centering}
\includegraphics[scale=0.32]{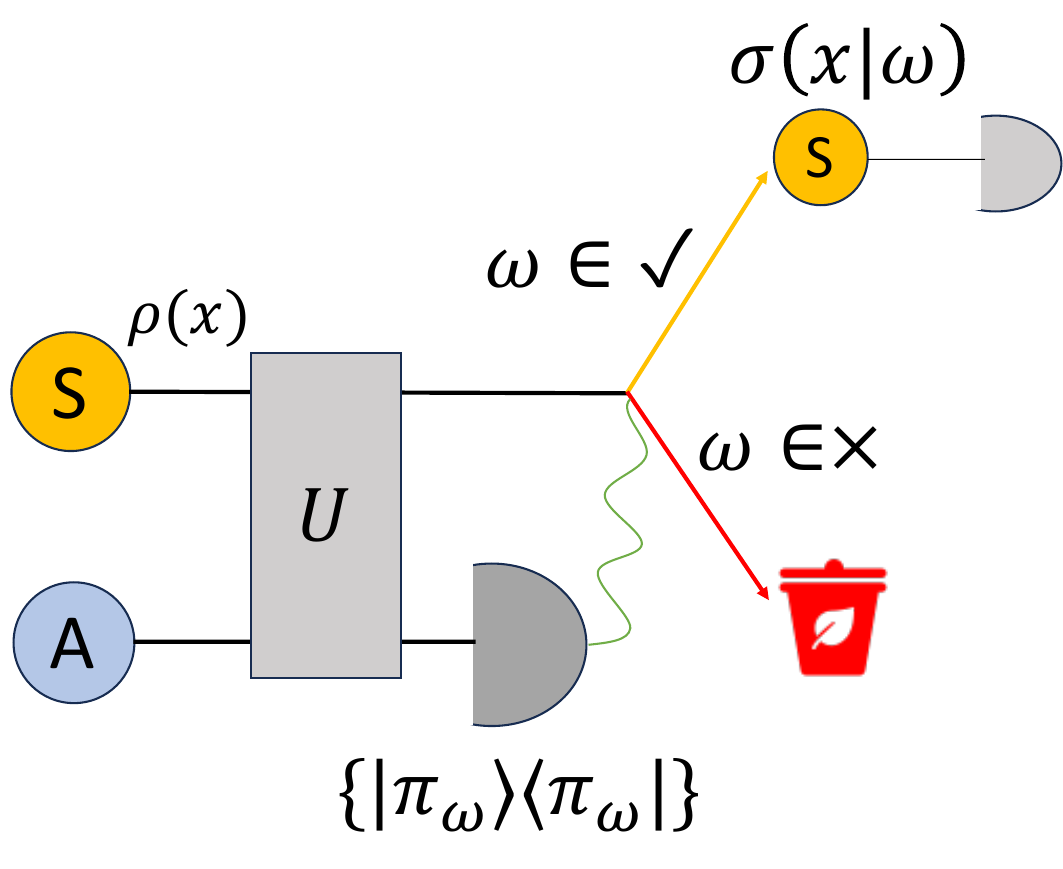}
\par\end{centering}
\caption{\protect\label{fig:PS-protocol}The postselection measurement can
be realized by entangling the probe system with an ancilla and a following
projective measurements on the ancilla. }
\end{figure}

\noindent\textbf{The postselection inequality. }As shown in Fig.~\ref{fig:PS-protocol},
a postselection measurement is usually realized by introducing an
ancilla and performing a projective measurement on the ancilla after
its coupling to the system. Effectively, this implements a postselection
measurement $\{M_{\omega}\}_{\omega\in\Omega}$ performed on the quantum
state of the probe system $\rho(x)$, where $x$ is the estimation
parameter, $\omega$ denotes each measurement outcome, and $\Omega$
denotes the set of all measurement outcomes. The QFI encoded in a
given state $\rho(x)$ is 
\begin{equation}
I^{Q}[\rho(x)]=\mathrm{Tr}[\rho(x)L^{2}(x)],\label{eq:IQ-def}
\end{equation}
where $L(x)$ is the symmetric logarithmic derivative (SLD)defined
as 
\begin{equation}
\partial_{x}\rho(x)=\frac{1}{2}[L(x)\rho(x)+\rho(x)L(x)].\label{eq:SLD-def}
\end{equation}

\noindent In standard quantum metrology, we require that the classical
Fisher information (CFI) encoded in the measurement statistics saturates
the QFI. In postselected quantum metrology, we would like the QFI
to be transferred to the postselected states~\citep{dressel2014colloquium,arvidsson-shukur2020quantum,yang2023theoryof,yang2024quantum}.
After performing the postselection measurement, the joint state of
the probe system and the ancilla becomes 
\begin{equation}
\sigma^{\text{SA}}(x)=\sum_{\omega\in\Omega}p(\omega|x)\sigma(x|\omega)\otimes\ket{\pi_{\omega}^{\mathrm{A}}}\bra{\pi_{\omega}^{\mathrm{A}}},
\end{equation}
where $p(\omega|x)=\text{Tr}[\rho(x)E_{\omega}]$ and $\sigma(x|\omega)=M_{\omega}\rho(x)M_{\omega}^{\dagger}/p(\omega|x)$.
The QFI corresponding to the state $\sigma^{\text{SA}}(x)$ is~\citep{yang2023theoryof,combes2014quantum},
\begin{equation}
I^{Q}[\sigma^{\text{SA}}(x)]=\sum_{\omega\in\Omega}I_{\omega}^{Q}[\sigma^{\text{SA}}(x)]\label{eq:I-sigma}
\end{equation}
where
\begin{equation}
I_{\omega}^{Q}[\sigma^{\text{SA}}(x)]\equiv I_{\text{\ensuremath{\omega}}}^{\text{cl}}[p(\omega\big|x)]+p(\omega\big|x)I^{Q}[\sigma(x|\omega)]\label{eq:I-omg}
\end{equation}
and 
\begin{equation}
I^{\text{cl}}[p(\omega\big|x)]\equiv\left[\partial_{x}p(\omega\big|x)\right]^{2}/p(\omega\big|x)\label{eq:Icl-omg}
\end{equation}
is the CFI associated with the measurement outcome $\omega$. We denote
the subset of desired outcomes and the subset of discarded outcomes
as $\checkmark$ and $\times$, respectively. 

\noindent Physically, since the measurement process is non-unitary,
the QFI cannot grow. This phenomenon is best characterized by the
following postselection inequality holds~\citep{shitara2016tradeoff,combes2014quantum,yang2023theoryof}
(see also Supplementary Note~\ref{SI:challenges}):
\begin{equation}
\sum_{\omega\in\checkmark}p(\omega\big|x)I^{Q}[\sigma(x|\omega)]\leq I^{Q}[\sigma^{\text{SA}}(x)]\leq I^{Q}[\rho(x)]\label{eq:PS-ineq}
\end{equation}
Refs.~\citep{arvidsson-shukur2020quantum} and \citep{yang2023theoryof}
further show that for pure states Eq.~(\ref{eq:PS-ineq}) is saturable,
which implies that postselection measurement can significantly reduce
the number of metrological samples without losing the precision. However,
realistic quantum systems are always subjected to decoherence. Extending
the previous theory for pure states to mixed states presents formidable
challenges, as it is hard to see when Eq.~(\ref{eq:PS-ineq}) holds
for mixed states, see Supplementary Note~\ref{SI:challenges} for
details

\medskip{}

\noindent\textbf{Lossless postselection of quasi-pure mixed states.} 

We consider the spectral decomposition of $\rho(x)$,
\begin{equation}
\rho(x)=\sum_{n=1}^{d_{r}}q_{n}(x)\ket{\varphi_{n}(x)}\bra{\varphi_{n}(x)},\label{eq:spec-decomp}
\end{equation}
where $\{\ket{\varphi_{n}(x)}\}_{n=1}^{d_{r}}$ is a set of orthonormal
basis, $d_{r}$ is the \textit{global} rank of $\rho(x)$ and $q_{n}(x)$
satisfies the global normalization condition 
\begin{equation}
\sum_{n=1}^{d_{r}}q_{n}(x)=1,\label{eq:global-constr}
\end{equation}
which holds for all values of $x$. Upon defining the global support
$\mathcal{H}_{r}\equiv\mathrm{span}\{\ket{\varphi_{n}(x)}\}_{n=1}^{d_{r}}$,
we can partition the Hilbert space to $\mathcal{H}=\mathcal{H}_{r}\oplus\mathcal{H}_{k}$,
where $\mathcal{H}_{k}$ is the global kernel of $\rho(x)$ with dimension
$d_{k}$. We note that while $q_{n}(x)$ may vanish locally at certain
value of $x$, the eigenvalues corresponding to the eigenstates in
$\mathcal{H}_{k}$ is identically vanishes, regardless of the values
of $x$. This observation leads to an elegant properties of the sensitivity
of $\rho(x)$ with respect to the infinitesimal changes of $x$, which
is relevant for the estimation of $x$, see ``Methods''. The global
rank should be distinguished from local rank. For example, consider
the state of two incoherent point sources, see Eq.~(\ref{eq:rho-super})
in what follows. This is a global rank-two state, though in the limit
$x\to0$ the state approaches a pure state. 

A more generic convex decomposition of $\rho(x)$ can be represented
by
\begin{equation}
\rho(x)=\sum_{n=1}^{w}p_{n}(x)\ket{\psi_{n}(x)}\bra{\psi_{n}(x)},\label{eq:Con-Decomp}
\end{equation}
where $p_{n}(x)$ is strictly positive and $\{\ket{\psi_{n}(x)}\}$
is a set of normal, linearly independent, but not necessarily orthogonal
vectors that spans $\mathcal{H}_{r}$ and $w\ge d_{r}$~\citep{fujiwara2008afibre}. 

Inspired by analysis on the amplification mechanism in the lossless
postselection for pure states (see ``Methods''), we consider the
following post-selection POVM element for mixed states,

\begin{equation}
E_{\checkmark}(x_{*})=\Pi_{k}(x_{*})+\lambda_{\checkmark}\Pi_{r}(x_{*}),\label{eq:QP-E-kernel}
\end{equation}
where $\lambda_{\checkmark}\in(0,1)$,$\Pi_{r}(x)$ and $\Pi_{k}(x)$
are the projectors to the global support $\mathcal{H}_{r}$ and the
global kernel $\mathcal{H}_{k}$ respectively, and $x_{*}$ represents
our prior knowledge. In local estimation approach, one always works
in the limit where $x_{*}$ is veery close to the true value of $x$.

We define a class of ``quasi-pure'' quantum states: $\rho(x)$ is
quasi-pure if $\Pi_{r}(x)\partial_{x}\rho(x)\Pi_{r}(x)=0$ or more
explicitly 
\begin{equation}
\braket{\varphi_{k}(x)|\partial_{x}\rho(x)|\varphi_{l}(x)}=0,\,\forall k,\,l=1,2,\cdots d_{r}.\label{eq:state-structure-spectral}
\end{equation}
Clearly, quasi-pure states include pure states as a special case.
Now we are in a position to state our main finding (see Supplementary
Note~\ref{SI:QP-PS-Proof} ):
\begin{thm}
\textup{\label{thm:QP-PS}Quasi-pure states can be lossless postselected
through the postselection measurement given by Eq.~(\ref{eq:QP-E-kernel})
in the limit $x_{*}\to x$.}
\end{thm}
One may ask: Given a representation of a quantum state, how to tell
whether it is quasi-pure or not? We now give several criteria. For
proofs, see Supplementary Note~\ref{SI:Criteria}.

\begin{obs} \label{obs:QP-spec}For a mixed $\rho(x)$ with the spectral
decomposition given by Eq.~(\ref{eq:spec-decomp}), it is quasi-pure
if and only if (i) The positive eigenvalues of $\rho(x)$ is insensitive
to the change of the estimation parameter, i.e.,
\begin{equation}
\partial_{k}q_{k}(x)=0,\,k=1,2,\,\cdots,d_{r}\label{eq:QP-cond1}
\end{equation}
(ii) The parameter derivative of $\ket{\varphi_{k}(x)}$ must be orthogonal
to an eigenstate $\ket{\varphi_{l}(x)}$ with different eigenvalues.
That is, at least one of the following condition should hold for $k,l=1,2,\cdots,d_{r}$
with $k\neq l$
\begin{subequations}
\label{eq:QP-cond2}

\begin{equation}
q_{k}(x)=q_{l}(x),\label{eq:QP-eq-prob}
\end{equation}

\begin{equation}
\braket{\partial_{x}\varphi_{k}(x)|\varphi_{l}(x)}=0.\label{eq:QP-deriv-ortho}
\end{equation}
\end{subequations}

\end{obs}

Observation~\ref{obs:QP-spec} can be easily seen upon inspecting
the expression of $\partial_{x}\rho(x)$ (see ``Methods''). The
second condition can be written in a more compact way. We recall the
conventional covariant derivative for pure states
\begin{equation}
\ket{\mathcal{D}_{x}\psi(x)}\equiv\ket{\partial_{x}\psi(x)}-\ket{\psi(x)}\braket{\psi(x)\big|\partial_{x}\psi(x)}.
\end{equation}
Here we introduce the generalized covariant derivative
\begin{equation}
\ket{\mathscr{D}_{x}\psi_{n}(x)}\equiv\ket{\partial_{x}\psi_{n}(x)}-\Pi_{n}(x)\ket{\partial_{x}\psi_{n}(x)},
\end{equation}
for quasi-pure states, where $\Pi_{n}(x)$ is the projector to the
degenerate subspace associated with the eigenvalue $q_{n}(x)$. Then
we obtain (see Supplementary Note~\ref{SI:Criteria}):

\begin{obs}\label{obs:GenCoDer}Condition (ii) in Observation~\ref{obs:QP-spec}
is equivalent to the orthogonality condition between the eigenstates
of $\rho(x)$ and their generalized covariant derivatives, i.e.,
\begin{equation}
\braket{\mathscr{D}_{x}\psi_{k}(x)|\psi_{l}(x)}=0,\,\forall k,l=1,2,\cdots,d_{r}.\label{eq:GenCoDer}
\end{equation}

\end{obs}

\begin{obs} \label{obs:QP-Conv}For a state represented by the convex
decomposition~(\ref{eq:Con-Decomp}), the state is quasi-pure if
and only if 
\begin{equation}
\braket{\psi_{k}(x)|\partial_{x}\rho(x)|\psi_{l}(x)}=0,\forall k,l=1,2,\cdots w.\label{eq:state-structure-convex}
\end{equation}
\end{obs}

Furthermore, quasi-pure states have several elegant mathematical properties
(see Supplementary Note~\ref{SI:Properties}):

\begin{obs} \label{obs:QP-SLD}The SLD of a mixed state satisfies
$\Pi_{r}(x)L(x)\Pi_{r}(x)=0$ and bear a simple form 
\begin{equation}
L(x)=\sum_{n}L_{n}(x),\label{eq:QP-SLD}
\end{equation}
where 
\begin{equation}
L_{n}(x)\equiv2\left(\ket{\mathscr{D}_{x}\psi_{n}(x)}\bra{\psi_{n}(x)}+\ket{\psi_{n}(x)}\bra{\mathscr{D}_{x}\psi_{n}(x)}\right).
\end{equation}
.

\end{obs}

\begin{obs} \label{obs:QP-QFI}The QFI for a quasi-pure mixed state
is 
\begin{equation}
I^{Q}[\rho(x)]=4\sum_{n=1}^{d_{r}}p_{n}(x)\braket{\mathscr{D}_{x}\varphi_{n}(x)|\mathscr{D}_{x}\varphi_{n}(x)}\label{eq:QP-QFI}
\end{equation}
 In particular, if the spectrum of a quasi-pure mixed states is non-degenerate,
\begin{equation}
I^{Q}[\rho(x)]=\sum_{n=1}^{d_{r}}p_{n}(x)I^{Q}[\ket{\varphi_{n}(x)}]\label{eq:QP-non-degenerate}
\end{equation}
which saturates the generalized convexity inequality of the QFI~\citep{alipour2015extended,yang2024quantum}:
\begin{equation}
I^{Q}[\rho(x)]\leq\sum_{n=1}^{w}\left(I^{\mathrm{cl}}[p_{n}(x)]+p_{n}(x)I^{Q}[\ket{\psi_{n}(x)}]\right),\label{eq:Gen-Conv}
\end{equation}
where $\{p_{n}(x),\,\ket{\psi_{n}(x)}\}$ is the convex decomposition
of $\rho(x)$ defined in Eq.~(\ref{eq:Con-Decomp}) and $I^{\mathrm{cl}}[p_{n}(x)]$
is defined as in Eq.~(\ref{eq:Icl-omg}).

\end{obs}

A few comments in order. First, if $q_{n}(x)$ is non-degenerate,
the generalized covariant derivative coincides with the conventional
covariant derivative and therefore$L_{n}(x)$ in the quasi-pure mixed
states reduces to the SLD for pure state $\ket{\psi_{n}(x)}$. Generally,
they bear very similar structure, up to the definition of the covariant
derivative. Using the definition of the SLD~(\ref{eq:SLD-def}) and
Eq.~(\ref{eq:QP-SLD}), we can express $\partial_{x}\rho(x)$ in
a more compact form in terms of the generalized covariant derivatives:
\begin{equation}
\partial_{x}\rho(x)=\sum_{n=1}^{d_{r}}q_{n}(x)\left[\ket{\mathscr{D}_{x}\varphi_{n}(x)}\bra{\varphi_{n}(x)}+\mathrm{h.c.}\right].
\end{equation}
It is then relevant to further decompose the global kernel space as
$\mathcal{H}_{k}=\mathcal{H}_{t}\oplus\mathcal{H}_{t}^{\perp}$, where$\mathcal{H}_{t}=\text{span}\{\ket{\mathscr{D}_{x}\varphi_{n}(x)}\subseteq\mathcal{H}_{k}$.
We denote the projectors to $\mathcal{H}_{t}$ and $\mathcal{H}_{t}^{\perp}$
as $\Pi_{t}(x)$ and $\Pi_{t}^{\perp}(x)$ respectively. Since $\Pi_{t}^{\perp}\rho(x)=\Pi_{t}^{\perp}\partial_{x}\rho(x)=0$,
$\Pi_{t}^{\perp}$ does not play a role in the postselection, i.e.,
it neither affect the post-measurement state nor the measurement statistics.
As a result, for quasi-pure states, Eq.~(\ref{eq:QP-E-kernel}) can
be modified as 
\begin{equation}
E_{\checkmark}(x_{*})=\Pi_{t}(x_{*})+\lambda_{\checkmark}\Pi_{r}(x_{*}),\,\lambda_{\checkmark}\in(0,1)\label{eq:QP-E-trans}
\end{equation}
Furthermore, similar with the case of pure states, it is possible
to deform Eq.~(\ref{eq:QP-E-trans}) into lossless post-selection
POVM element with multiple desired outcomes, 
\begin{equation}
E_{\omega}(x_{*})=\mu_{\omega}\Pi_{t}(x_{*})+\lambda_{\omega}\Pi_{r}(x_{*}),
\end{equation}
where $\omega\in\checkmark$ and $\sum_{\omega\in\checkmark}\mu_{\omega}=1$.
Any positive operator whose support fully lies in $\mathcal{H}_{t}^{\perp}$
can be also added to $E_{\omega}(x_{*})$, without incurring any loss
of QFI. The special case would be adding $\Pi_{t}^{\perp}$ to (\ref{eq:QP-E-trans}),
then it becomes 
\begin{equation}
E_{\omega}(x_{*})=\mu_{\omega}\Pi_{k}(x_{*})+\lambda_{\omega}\Pi_{r}(x_{*})\label{eq:QP-E-kernel-multi}
\end{equation}

Next, the definition of the quasi-pure structure straightforwardly
generalizes to the multi-parameter case upon imposing Eq.~(\ref{eq:state-structure-spectral})
for all the estimation parameters. Then, the multi-parameter version
of lossless postselection POVM can be read off directly from Eq.~(\ref{eq:QP-E-kernel})
and Eq.~(\ref{eq:QP-E-kernel-multi}), respectively:
\begin{equation}
E_{\checkmark}(\bm{x}_{*})=\mathbb{I}+(\lambda_{\checkmark}-1)\Pi_{r}(\bm{x}_{*}),\;E_{\omega}(\bm{x}_{*})=\mu_{\omega}\mathbb{I}+(\lambda_{\omega}-\mu_{\omega})\Pi_{r}(\bm{x}_{*}),
\end{equation}
where $\omega\in\checkmark$, $\lambda_{\checkmark}$, $\lambda_{\omega}$
and $\mu_{\omega}$ are defined previously. Furthermore, we have the
following observation for multi-parameter quasi-pure states, which
again confirms its pure-state like properties (see Supplementary Note~\ref{SI:Properties}): 

\begin{obs}\label{obs:QP-PCC}For multi-parameter quasi-pure states,
the compatibility condition at the single-copy level, known as the
partial commutativity condition~\citep{yang2019optimal}
\begin{equation}
\braket{\psi_{k}(\bm{x})|[L(x_{i}),\,L(x_{j})]|\psi_{l}(\bm{x})}=0
\end{equation}
 reduces to
\begin{equation}
\braket{\mathscr{D}_{x_{i}}\psi_{k}(\bm{x})|\mathscr{D}_{x_{j}}\psi_{l}(\bm{x})}=\braket{\mathscr{D}_{x_{j}}\psi_{k}(\bm{x})|\mathscr{D}_{x_{i}}\psi_{l}(\bm{x})}\label{eq:QP-PCC}
\end{equation}
for all $k,l=1,2,\cdots d_{r}$ and all pair of estimation parameters
$x_{i}$ and $x_{j}$, where $L(x_{i})$ and $\mathscr{D}_{x_{i}}$
are the SLD and the generalized co-variant derivative with respect
to the estimation parameter $x_{i}$, respectively.

\end{obs}

Note that for pure states, Eq.~(\ref{eq:QP-PCC}) immediately reduces
to the weak commutativity condition~\citep{matsumoto2002anew,pezz`e2017optimal}.
We conclude this section by noting that quasi-pure structures only
exist for the \textit{global} rank-deficient states. When $\rho(x)$
is full-rank globally, protocols to enlarge the dimension of the Hilbert
space is necessary so that the state becomes rank-deficient in the
enlarged system. This can be achieved via introducing an ancilla or
exploring hidden levels, where the Hilbert space is enlarged through
tensor product or direct sum respectively. 

\medskip{}

\noindent\textbf{Applications}

Despite the elegant mathematical structure of quasi-pure states, one
may wonder how they are relevant to the practical quantum metrological
problems. In this section, we give three examples. In the first example,
we consider postselection in quantum imaging. The state is approximately
quasi-pure locally in some neighborhood of $x=0$. In the second example,
we consider unitary estimation and the state is globally quasi-pure
for all values $x$. In the last example, we consider unitary estimation
with the assistance of ancilla. We propose a universal protocol to
engineer the quasi-pure structures globally regardless of the values
of $x$. 

It should be noted that even though the quasi-pure structure can be
global, the lossless postselection measurement discussed throughout
this work is always local, i.e. it requires refined prior knowledge
of the estimation. For all the examples, we consider binary postselection
and take $M_{\checkmark}(x_{*})=\Pi_{t}(x_{*})+\sqrt{\lambda_{\checkmark}}\Pi_{r}(x_{*})$.
We introduce the following figures of merit
\begin{equation}
\varepsilon_{0}(x)\equiv\|\sigma(x|\checkmark)-\rho(x)\|,\,\varepsilon_{1}(x)\equiv\|\partial_{x}\sigma(x|\checkmark)-\partial_{x}\rho(x)/\sqrt{\lambda_{\checkmark}}\|,\label{eq:vareps-def}
\end{equation}
to characterize the performance of the postselection measurement.
If $\rho(x)$ is quasi-pure and $x_{*}$ is close to the true value
of $x$, then $\varepsilon_{0}(x)$ and $\varepsilon_{1}(x)$ will
be very close to zero, indicating the the postselection measurement
is lossless. 

\medskip{}

\noindent\textbf{Application 1: Local quasi-pure states in postselected
quantum imaging}

As a first example, we consider the superresolution imaging of two
incoherent point sources with a Gaussian point-spread function~\citep{tsang2016quantum}.
The quantum state of the two point sources is described by globally
rank-two density operator as follows:
\begin{equation}
\rho(x)=q\ket{\psi_{+}(x)}\bra{\psi_{+}(x)}+(1-q)\ket{\psi_{-}(x)}\bra{\psi_{-}(x)}\label{eq:rho-super}
\end{equation}
where $\ket{\psi_{\pm}(x)}=e^{-i\hat{P}\left(\pm\frac{x}{2}\right)}\ket{\psi_{0}}$,
$\braket{u|\psi_{0}}=\left(\frac{1}{2\pi\sigma^{2}}\right)^{1/4}e^{-\frac{u^{2}}{4\sigma^{2}}}$
and $\hat{P}$ is the momentum operator defined as $\braket{u|\hat{P}|\psi}=-\text{i}\partial_{u}\braket{u|\psi}$~\footnote{It should not be confused that here $u$ denotes the position while
$x$ denotes the estimation parameter. }. Since we focus on single-parameter estimation and study the fundamental
limits of post-selection, we assume the intensities of the sources
are known. In this case, it can shown that $I^{Q}[\rho(x)]=1/(4\sigma^{2})$,
independent of the values of the source intensities and the values
of $x$~\citep{vrehavcek2017multiparameter,vrehavcek2018optimal}.

\begin{figure}
\begin{centering}
\includegraphics[scale=0.42]{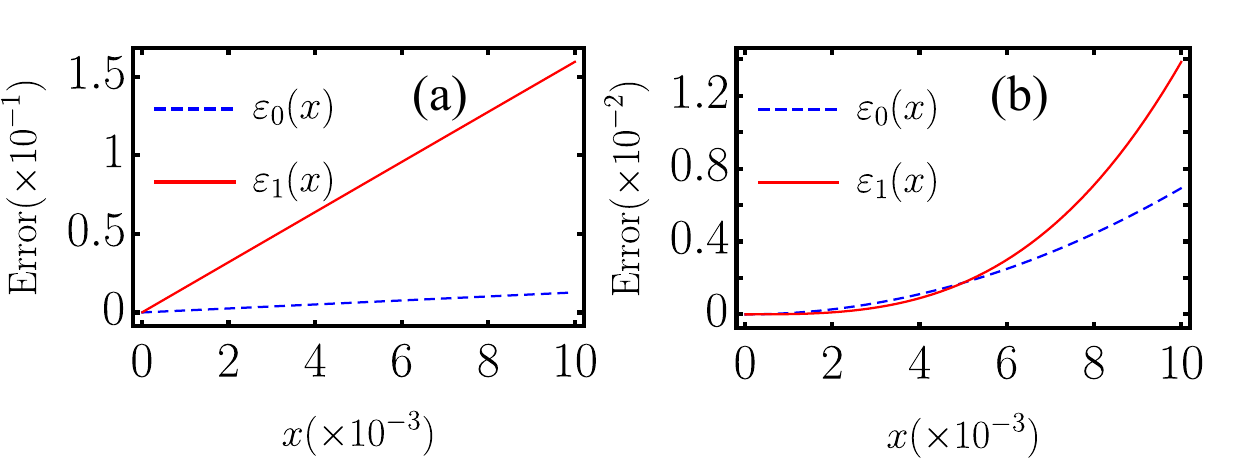}
\par\end{centering}
\caption{\protect\label{fig:Errors}The errors of the post-selection measurements
Eq.~(\ref{eq:QP-E-kernel}) and Eq.~(\ref{eq:E-2qu}) for (a) superresolution
and (b) two-qubit example in the limit $x\to0$, respectively. The
norms used in the numerical calculation of Eq.~(\ref{eq:vareps-def})
are the $L^{2}$-norm and matrix-$2$ norm for (a) and (b), respectively.
Values of parameters: (a) $\lambda=10^{-2}$, $\sigma=1$, $q=0.3$.
(b)$\lambda=10^{-4}$, $q=0.3$. }
\end{figure}

Intuitively, in the Rayleigh limit $x\to0$, $\ket{\psi_{\pm}(x)}\to\ket{\psi_{0}}$
and the state~(\ref{eq:rho-super}) approaches a pure state. Therefore
we expect in the neighborhood of $x=0$, the state is \textit{approximately}
quasi-pure. In this limit, both $\mathcal{H}_{t}$ and $\mathcal{H}_{r}$
become the rank$-1$ subspaces with bases $\ket{\psi_{1}}$ and $\ket{\psi_{0}}$,
respectively, where $\langle u|\psi_{1}\rangle=u\braket{u|\psi_{0}}/\sigma$
is the first-order Hermite-Gaussian function.

More precisely, since Eq.~(\ref{eq:rho-super}) is a convex decomposition,
we can directly verify this intuition using Observation~\ref{obs:QP-Conv}.
It can be calculated that for all values of $x$, 
\begin{align*}
\braket{\psi_{+}(x)|\partial_{x}\rho(x)|\psi_{+}(x)} & =2q\text{i}\langle\hat{P}e^{-\text{i}\hat{P}x}\rangle\langle e^{\text{i}\hat{P}x}\rangle,\\
\braket{\psi_{-}(x)|\partial_{x}\rho(x)|\psi_{-}(x)} & =2(1-q)\text{i}\langle\hat{P}e^{-\text{i}\hat{P}x}\rangle\langle e^{\text{i}\hat{P}x}\rangle,\\
\braket{\psi_{+}(x)|\partial_{x}\rho(x)|\psi_{-}(x)} & =\text{i}\langle\hat{P}e^{-\text{i}\hat{P}x}\rangle,
\end{align*}
where the average is taken over the state $\ket{\psi_{0}}$. It is
clear that Eq.~(\ref{eq:state-structure-convex}) is satisfied in
the limit $x\to0$ thanks to $\langle\hat{P}\rangle=0$.

Therefore, in the Rayleigh limit, we can use the post-selection POVM
element 
\begin{equation}
E_{\checkmark}=\ket{\psi_{1}}\bra{\psi_{1}}+\lambda_{\checkmark}\ket{\psi_{0}}\bra{\psi_{0}}\label{eq:E-Super}
\end{equation}
to reduce the number of detected photons. The performance of this
Eq.~(\ref{eq:E-Super}) is shown in Fig.~\ref{fig:Errors}(a). Under
the $L^{2}$-norm defined as $\|f\|=\sqrt{\int_{-\infty}^{\infty}|f(u,u^{\prime})|^{2}dudu^{\prime}}$,
it can be analytically calculated straightforwardly that to the leading
order of $x$, $\varepsilon_{0}(x)=|2q-1|(1-\sqrt{\lambda})x/(2\sqrt{2}\sqrt{\lambda}\sigma)$
and $\varepsilon_{1}(x)=\sqrt{2-4\sqrt{\lambda}+3\lambda}x/(8\lambda\sigma^{2})$.

It is straightforward to generalize the arguments here to the case
of more realistic Zernike point spread function~\citep{yang2019optimal}
or the estimation of the longitudinal separation ~\citep{zhou2019quantumlimited}.

We conclude by discussing the practical relevance of applying the
postselection protocol in optical imaging. In realistic scenarios,
photon number and time are also valuable resources. For fixed amount
of time, due to detector saturation, the resolution cannot increase
indefinitely, as the number of photons increases. Using ideas from
postselected metrology for optical imaging brings the advantage of
preserving the QFI significantly while avoiding detector saturation,
as in the case of weak value amplification~\citep{harris2017weakvalue,xu2020approaching}.

We can analyze the advantage more precisely in the context imaging
two point sources with the lossless POVM~(\ref{eq:E-Super}) in the
Rayleigh limit $x\to0$. We denote the total number of photons and
time for imaging as $N$ and $T$, respectively. We consider a detector
model that can detect at most $N_{0}$-photons within a given time
slot $\Delta t$, i.e., the detection rate is $\gamma=N_{0}/\Delta t$.
In the case of illumination with weak source, the average illumination
rate $N/T$ is much smaller than $\gamma$, postselection measurement
offer no advantage over the standard postselection free case as the
detector does not saturate. As a result, we expect that postselection
is advantageous when $N\gg N_{\text{cr}}=T\gamma$. In this case,
one can tune the postselection probability $\lambda_{\checkmark}$
such that $\lambda_{\checkmark}<T\gamma/N$.\medskip{}

\noindent\textbf{Application 2: Global quasi-pure structure in unitary
estimation without ancilla}

\noindent We consider unitary estimation with mixed initial state,
i.e. $\rho(x)=U(x)\rho_{\mathrm{i}}U^{\dagger}(x)$. We denote the
spectral decomposition of the initial $\rho_{\mathrm{i}}$ as $\rho_{\mathrm{i}}=\sum_{n}q_{n\mathrm{i}}\ket{\phi_{n\mathrm{i}}}\bra{\phi_{n\mathrm{i}}}$,
where $q_{n\mathrm{i}}>0$. From Observation~\ref{obs:QP-spec},
we see that $\rho(x)$ is quasi-pure if and only if one of the following
two condition. 
\begin{align}
q_{k\mathrm{i}} & =q_{l\mathrm{i}}\label{eq:init-eq-q}\\
\langle\phi_{k\mathrm{i}}|H(x)|\phi_{l\mathrm{i}}\rangle & =0\label{eq:init-ortho}
\end{align}
where $H(x)=\text{i}\partial_{x}U^{\dagger}(x)U(x)$. Given any distinct
pair of $k,l$, we know at least one the following conditions must
hold. Using Eq.~(\ref{eq:QP-QFI}), it can be found that 
\begin{equation}
I^{Q}[\rho(x)]=4\left[\sum_{k}q_{k\mathrm{i}}\text{Var}[H(x)]_{\ket{\phi_{k\mathrm{i}}}}\!-\!\sum_{k,l,\,k\neq l,q_{l\mathrm{i}}=q_{k\mathrm{i}}}q_{k\mathrm{i}}|\langle\phi_{k\mathrm{i}}|H(x)|\phi_{l\mathrm{i}}\rangle|^{2}\right]
\end{equation}

For the estimation of two-qubit unitary $U(x)=e^{-\text{i}\sigma_{x}^{(1)}\sigma_{x}^{(2)}x}$,
the optimal pure initial state for estimating $x$ can be $\ket{\phi_{1\mathrm{i}}}=\ket{00}$
and $\ket{\phi_{2\mathrm{i}}}=\ket{01}$. Suppose there is uncertainty
in preparing the second qubit in the computation basis so that the
initial state becomes mixed, i.e. $\rho_{\mathrm{i}}=q_{1\mathrm{i}}\ket{\phi_{1}}\bra{\phi_{1}}+(1-q_{1\mathrm{i}})\ket{\phi_{2}}\bra{\phi_{2}}$.
It can be readily calculated that Eq.~(\ref{eq:init-ortho}) is satisfied,
i.e., $\braket{\phi_{1}|\sigma_{x}^{(1)}\sigma_{x}^{(2)}|\phi_{2}}=0$
and $I^{Q}[\rho(x)]=I^{Q}[\ket{\varphi_{1}(x)}]=I^{Q}[\ket{\varphi_{2}(x)}]=4$,
where $\ket{\varphi_{n}(x)}=U(x)\ket{\phi_{n\mathrm{i}}}$. Thus $\rho(x)$
is universally quasi-pure for all values of $x$.

In ultra-sensitive estimation where $x\to0$ , it can be readily found
that $\mathcal{H}_{t}=\text{span}\{\ket{11},\,\ket{10}\}$ and $\mathcal{H}_{r}=\text{span}\{\ket{00},\ket{01}\}$.
Then 
\begin{equation}
E_{\checkmark}=\ket{1^{(1)}}\bra{1^{(1)}}+\lambda_{\checkmark}\ket{0^{(1)}}\bra{0^{(1)}}\label{eq:E-2qu}
\end{equation}
becomes the post-selection measurement only on the first qubit, reminiscent
of the weak value amplification. The performance of Eq.~(\ref{eq:E-2qu})
is shown in Fig.~\ref{fig:Errors}(b).

\medskip{}

\noindent\textbf{Application 3: Engineering global quasi-pure states
in unitary estimation with ancilla}

\begin{figure}
\begin{centering}
\includegraphics[scale=0.28]{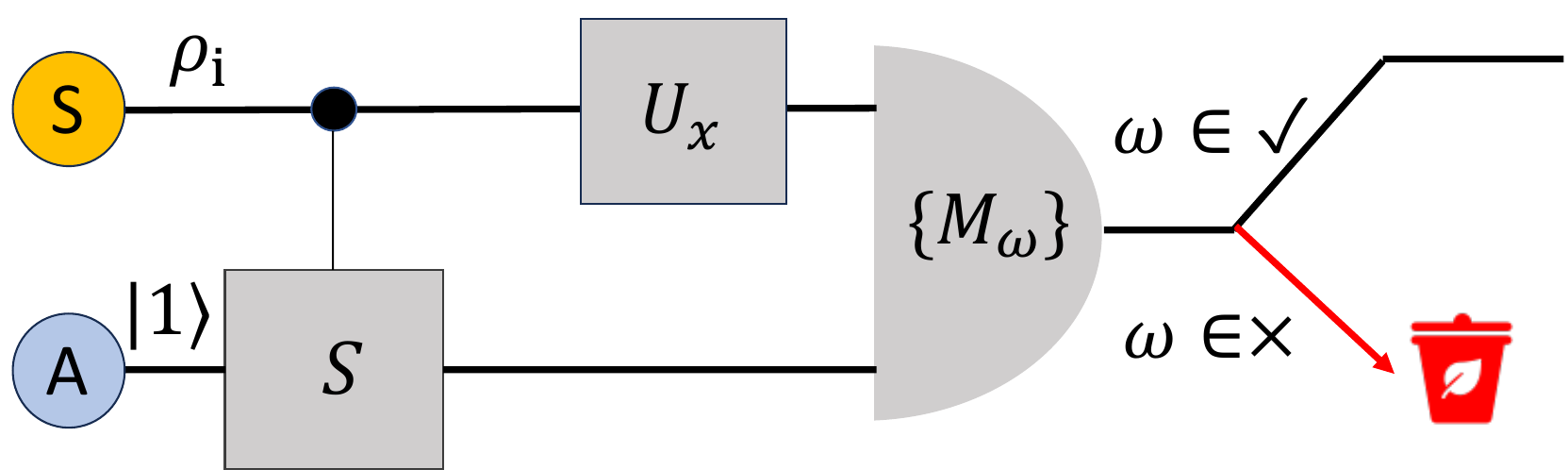}
\par\end{centering}
\caption{\protect\label{fig:universal-protocol}The universal protocol to create
the quasi-pure structure by introducing ancilla. The dimension of
the Hilbert space of the ancilla is at least $d_{r}$. The initial
state of the ancilla is prepared in $\ket{1}$. After the control
sum gate, the state become Eq.~(\ref{eq:sig-init}) in the main text.}

\end{figure}

\noindent Having discussed the particular cases, let us now present
a universal protocol, as shown in Fig.~\ref{fig:universal-protocol},
which can engineer the quasi-pure structures, regardless of the values
of the estimation parameter. 

The crucial procedure is to introduce an ancilla and create classical
correlations between the probe system and the ancilla before the unitary
encoding. The classical correlated state is 
\begin{equation}
\sigma_{\mathrm{i}}=\sum_{n=1}^{d_{r}}q_{n\mathrm{i}}\ket{\phi_{n\mathrm{i}}}\bra{\phi_{n\mathrm{i}}}\otimes\ket{n}\bra{n},\label{eq:sig-init}
\end{equation}
where $\{\ket{n}\}_{n=1}^{d_{r}}$ is a set of orthonormal states
in the ancillary Hilbert space $\mathcal{H}_{\mathrm{A}}$. Such a
state can be created via the control sum gate CS as shown \ref{fig:universal-protocol},
which $\mathrm{CS}\ket{\phi_{n\mathrm{i}}}\ket{k}\to\ket{\phi_{n\mathrm{i}}}\ket{n+k-1}$.
It should be noted such an ancilla should be distinguished from the
ancilla that is used in implementing the postselection measurement
in Fig.~\ref{fig:PS-protocol}. 

Then after unitary encoding, the state of the composite system becomes
\begin{equation}
\sigma(x)=\sum_{n=1}^{d_{r}}q_{n\mathrm{i}}\ket{\varphi_{n}(x)}\bra{\varphi_{n}(x)}\otimes\ket{n}\bra{n}
\end{equation}
Since unitary encoding does not change the eigenvalues of $\sigma_{\mathrm{i}}$
so Eq.~(\ref{eq:QP-eq-prob}) holds. Furthermore, thanks to the classical
correlation with the ancilla, Eq.~(\ref{eq:QP-deriv-ortho}) is also
satisfied. Therefore, $\sigma(x)$ is a quasi-pure state, regardless
of the values of $x$.

The lossless postselection can be analyzed in a similar manner with
previous two examples and hence is omitted here. For the multi-parameter
case, the partial commutativity condition for quasi-pure state Eq.~(\ref{eq:QP-PCC})
in this case becomes $\mathrm{Im}\braket{\mathscr{D}_{x_{i}}\varphi_{n}(\bm{x})|\mathscr{D}_{x_{j}}\varphi_{n}(\bm{x})}=0$
for all pairs of $x_{i}$, $x_{j}$ and $n=1,2,\cdots d_{r}$. 

\medskip{}

\noindent{\large\textbf{Discussion}}{\large\par}

In this work, we give an intuitive understanding of the amplification
effect in lossless postselected quantum metrology for pure states:
post-selection amplifies the parametric derivative of the density
operator while preserving the original state. Based on this intuition,
we develop a general theory for the lossless postselection of a broad
class of mixed states, dubbed ``quasi-pure'' states. These states
possess structures analogous to the pure states when it comes to lossless
postselection. 

To demonstrate the broad applicability of our findings, we show that
the quasi-pure structure appears in postselected superresolution imaging
of two incoherent point sources and in unitary estimation with mixed
initial states. Furthermore, we propose a simple universal protocol
to create the quasi-pure structure in unitary estimation only using
classical correlations with ancillary systems. 

In the future, it is promising to engineer quasi-pure states and exploit
postselection to suppress technical noise~\citep{jordan2014technical},
perform efficient distributed quantum sensing~\citep{pezze_advances_2025}
and quantum imaging~\citep{tsang_resolving_2019}. On the fundamental
level, since the quasi-pure states behave like pure states, they may
provide insights into quantum information geometric problems of mixed
states, including the tightness of quantum speed limits for mixed
states~\citep{mandelstam1945theuncertainty,nicholson_timeinformation_2020,anandan1990geometry,delcampo2013quantum},
the construction of optimal measurements when a mixed state satisfies
the partial commutativity condition~\citep{yang2019optimal,horodecki2022fiveopen}. 

\medskip{}

\noindent{\large\textbf{Methods}}{\large\par}

\noindent{\small\textbf{Global rank, support and kernel of $\rho(x)$.
}}{\small We can characterize several nice properties of $\rho(x)$
and $\partial_{x}\rho(x)$ using the notions of global rank, support
and kernel.}{\small\par}

\noindent{\small\begin{obs}\label{obs:eign-ker}The eigenvalues
with eigenstates belongs to the global kernel $\mathcal{H}_{k}$ must
be identically zero for all values of $x$.\end{obs}}{\small\par}
\begin{proof}
{\small We prove by contradiction. We denote the eigenvalues with eigenstates
in $\mathcal{H}_{k}$ as $\tilde{q}_{m}(x)$ with $m=1,2,\cdots,d_{k}$.
If there exists $m$ and $y$ such that $\tilde{q}_{m}(y)>0$. Then
$\sum_{n=1}^{d_{r}}q_{n}(y)$ must be strictly less than one, which
is in contradiction with the global normalization constraint. }{\small\par}
\end{proof}
{\small With Observation~(\ref{obs:eign-ker}), it is straightforward
to calculate
\begin{align}
\partial_{x}\rho(x) & =\sum_{n=1}^{d_{r}}\partial_{x}q_{n}(x)\ket{\varphi_{n}(x)}\bra{\varphi_{n}(x)}\nonumber \\
 & +\sum_{n=1}^{d_{r}}q_{n}(x)\ket{\partial_{x}\varphi_{n}(x)}\bra{\varphi_{n}(x)}+\sum_{n=1}^{d_{r}}q_{n}(x)\ket{\varphi_{n}(x)}\bra{\partial_{x}\varphi_{n}(x)}.\label{eq:drho-dx}
\end{align}
It should be noted that had we focused on the local support and kernel,
there would be contributions to $\partial_{x}\rho(x)$ from the zero
eigenvalues in the local kernel. This is because locally the eigenvalues
of $\rho(x)$ can vanish, its derivative with respect to $x$ does
not necessarily vanish. From Eq.~(\ref{eq:drho-dx}) that 
\begin{equation}
\Pi_{k}(x)\partial_{x}\rho(x)\Pi_{k}(x)=0
\end{equation}
is an identity holding for all density operators.}{\small\par}

\medskip{}

\noindent{\small\textbf{Pure-state inspired approach to the postselection
of mixed states.}}{\small{} We show that by leveraging the physics of
lossless postselection of pure states, we naturally arrive at the
quasi-pure structure. To this end, let us revisit the post-selection
of pure states discussed in Ref.~\citep{yang2023theoryof}. For the
sake of simplicity, at the moment, we shall focus on binary post-selection,
where $\Omega=\{\checkmark,\,\times\}$. We consider a postselection
POVM element of the following form:
\begin{equation}
E_{\checkmark}(x_{*})=\ket{\psi^{\perp}(x_{*})}\bra{\psi^{\perp}(x_{*})}+\lambda_{\checkmark}\ket{\psi(x_{*})}\bra{\psi(x_{*})},\label{eq:E-pure}
\end{equation}
where $\lambda_{\checkmark}\in(0,1)$, $x_{*}$ represents our prior
knowledge of the estimation parameter, $\ket{\psi^{\perp}(x)}\equiv\ket{\mathcal{D}_{x}\psi(x)}/\|\ket{\mathcal{D}_{x}\psi(x)}\|$
and $\|\cdot\|$ denotes the norm of a vector. In the limit $x_{*}\to x$,
$E_{\checkmark}(x_{*})$ becomes exact lossless. The post-selection
measurement operator is
\begin{equation}
M_{\checkmark}(x_{*})=U_{\checkmark}(x_{*})\sqrt{E_{\checkmark}(x_{*})},
\end{equation}
where $U_{\checkmark}(x_{*})$ is some unitary operator that may or
may not depend on $x_{*}$. Upon taking $U_{\checkmark}=\mathbb{I}$,
it is calculated in Ref.\citep{yang2023theoryof} that in the limit
$x_{*}\to x$, $\ket{\psi(x|\checkmark)}=\ket{\psi(x)}$ and
\begin{equation}
\ket{\partial_{x}\psi(x|\checkmark)}=\frac{1}{\sqrt{\lambda_{\checkmark}}}\ket{\partial_{x}\psi(x)}+\left(\frac{1}{\sqrt{\lambda_{\checkmark}}}-1\right)\ket{\psi(x)}\braket{\psi(x)|\partial_{x}\psi(x)}\label{eq:state-derivative}
\end{equation}
Clearly, one can see qualitatively on the amplification mechanism
of the post-selected QFI in the lossless postselection scheme: The
smaller the success probability is, the larger the prefactor $1/\sqrt{\lambda_{\checkmark}}$
on the first term of the r.h.s of Eq.~(\ref{eq:state-derivative})
becomes. Here, we emphasize such an intuition is not quantitative
as the role of the second term on the r.h.s. of Eq.~(\ref{eq:state-derivative})
is not clear. Furthermore, such an intuition does not necessarily
generalize to mixed states. However, in terms of the density operator,
we observe that}{\small\par}

{\small{}
\begin{subequations}
{\small\label{eq:DO-derivative}}{\small\par}

{\small
\begin{equation}
\sigma(x|\checkmark)=\rho(x),\label{eq:zeroth-order}
\end{equation}
}{\small\par}

{\small
\begin{equation}
\partial_{x}\sigma(x|\checkmark)=\frac{1}{\sqrt{\lambda_{\checkmark}}}\partial_{x}\rho(x).\label{eq:first-order}
\end{equation}
}{\small\par}
\end{subequations}
}{\small According to Eq.~(\ref{eq:SLD-def}), Eq.~(\ref{eq:DO-derivative})
immediately implies
\begin{equation}
L(x|\checkmark)=\frac{1}{\sqrt{\lambda_{\checkmark}}}L(x),
\end{equation}
where $L(x|\checkmark)$ and $L(x)$ are the SLDs for $\sigma(x|\checkmark)$
and $\rho(x)$, respectively. It follows from the generic definition
of QFI~(\ref{eq:IQ-def}) that 
\begin{equation}
I^{Q}[\sigma(x|\checkmark)]=\frac{1}{\lambda_{\checkmark}}I^{Q}[\rho(x)].\label{eq:IQ-amp}
\end{equation}
Here, one can clearly see that the intuition responsible for the amplification
effect of the post-selected QFI is due to the amplification of the
SLD. Apparently, it follows from Eq.~(\ref{eq:IQ-amp}) that the
bound~(\ref{eq:PS-ineq}) is saturated, implying the post-selection
measurement is lossless. }{\small\par}

{\small We would like to emphasize that Eqs.~(\ref{eq:DO-derivative}-\ref{eq:IQ-amp})
unveil the physics of the lossless post-selection measurement. A natural
extension of the postselection measurement~(\ref{eq:E-pure}) for
pure states is Eq.~(\ref{eq:QP-E-kernel}). We consider Eq.~(\ref{eq:QP-E-kernel})
and impose Eqs.~(\ref{eq:zeroth-order},~\ref{eq:first-order})
as additional constraints.}{\small\par}

{\small It can be shown that Eq.~(\ref{eq:zeroth-order}) is satisfied
while Eq.~(\ref{eq:first-order}) leads to the quasi-pure condition
$\Pi_{r}(x)\partial_{x}\rho(x)\Pi_{r}(x)=0.$ Since this condition
is inspired by the lossless postselection of pure states, we shall
refer to quantum states that satisfy this condition }{\small\textit{quasi-pure}}{\small{}
states. }{\small\par}

\medskip{}

\noindent{\large\textbf{Data Availability}}{\large\par}

\noindent All data relevant to this study are available from the corresponding
author upon request.

\medskip{}

\noindent{\large\textbf{Code Availability}}{\large\par}

\noindent Source codes of the plots are available from the corresponding
author upon request.

\medskip{}

\noindent{\large\textbf{References}}{\large\par}

\bibliographystyle{naturemag}
\bibliography{Post-Selected-Metrology}

\medskip{}

\noindent{\large\textbf{Acknowledgement}}{\large\par}

\noindent This work was supported by Zhejiang University startup grants
and the Wallenberg Initiative on Networks and Quantum Information
(WINQ) program. 

\medskip{}

\noindent{\large\textbf{Author contributions}}{\large\par}

\noindent J.Y. initiated the project and performed analytical and
numerical analysis.\medskip{}

\noindent{\large\textbf{Competing interests}}{\large\par}

\noindent The author declare no competing interests

\medskip{}

\noindent{\large\textbf{Additional Information}}{\large\par}

\noindent\textbf{Supplemental Information} The online version contains
supplementary material available at XXX.

\clearpage\newpage\setcounter{equation}{0} \setcounter{section}{0}\setcounter{figure}{0}
\setcounter{subsection}{0}\newcounter{suppnote} 
\global\long\def\theequation{S\arabic{equation}}%

\global\long\def\thefigure{S\arabic{figure}}%

\onecolumngrid \setcounter{enumiv}{0} 
\begin{center}
{\large\textbf{Supplementary Information}}{\large\par}
\par\end{center}

\tableofcontents{}

\refstepcounter{suppnote}

\section*{Supplementary Note 1: Challenges in the lossless postselection theory
for mixed states}

\label{SI:challenges}

Before we discuss the challenges for the lossless postselection of
mixed states, let us consider pure states, i.e., $\rho(x)=\ket{\psi(x)}\bra{\psi(x)}$,
Ref.~\citep{yang2023theoryof} shows that
\begin{equation}
p(\omega\big|x)I^{Q}[\sigma(x|\omega)]\leq I_{\omega}^{Q}[\sigma^{\text{SA}}(x)]\leq I_{\omega}^{Q}[\ket{\psi(x)}],\label{eq:meas-ineq-omg}
\end{equation}
where 
\begin{equation}
I_{\omega}^{Q}[\ket{\psi(x)}]=4\braket{\mathcal{D}_{x}\psi(x)\big|E_{\omega}\big|\mathcal{D}_{x}\psi(x)}
\end{equation}
is the covariant derivative~\citep{braunstein1994statistical,yang2019optimal}.
Summing over $\omega$, one obtains Eq.~(\ref{eq:PS-ineq}). For
pure states, Ref.~\citep{yang2023theoryof} systematically discussed
the conditions and postselection measurements that saturate Eq.~(\ref{eq:meas-ineq-omg})
by tracking its derivation.

For mixed states, Eq.~(\ref{eq:PS-ineq}) still holds. This can be
seen by constructing an effective completely positive and trace-preserving
(CPTP) map corresponding to the measurement process~\citep{shitara2016tradeoff}
and exploiting the monotonicity property of the QFI under a CPTP map~\citep{petz2011introduction}.
However, in the case of postselection of mixed states it is difficult
to see when Eq.~(\ref{eq:PS-ineq}) saturates due to technical difficulties.

\begin{figure}
\begin{centering}
\includegraphics[scale=0.32]{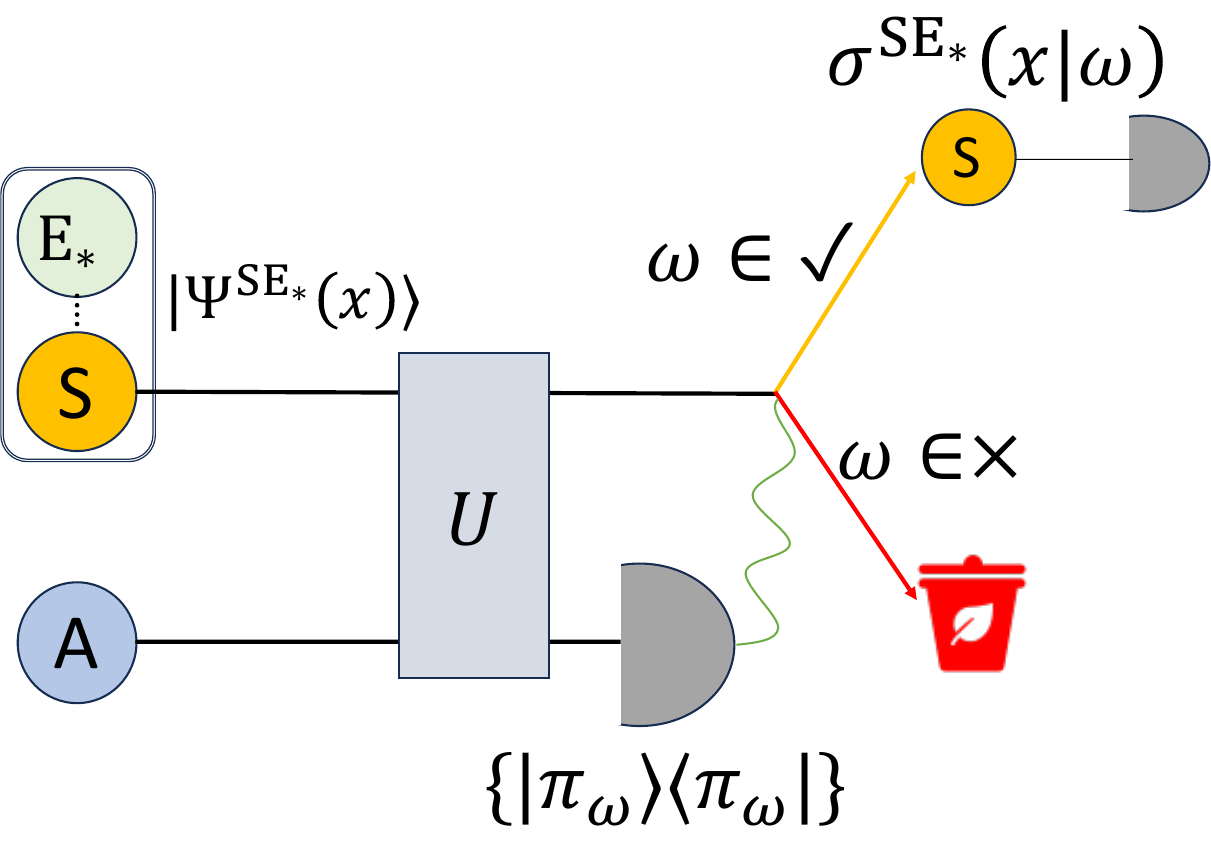}
\par\end{centering}
\caption{\protect\label{fig:purification} The postselection of a probe system
in mixed states can be artificially described as the postselection
on the joint system consisting of the probee system and the optimal
environment. An ancilla is introduced to implement the postselection
measurement.}
\end{figure}

To see in a more clear manner, let us now prove Eq.~(\ref{eq:PS-ineq})
using the idea of purification. According to the Uhlmann's theorem~\citep{uhlmann1976thetransition,escher2011general},
\begin{equation}
I^{Q}[\rho(x)]\leq I^{Q}(\ket{\Psi^{\text{SE}}(x)}),
\end{equation}
where $\ket{\Psi^{\text{SE}}(x)}$ is a purification of $\rho(x)$
and the upper bound can be saturated by an optimal environment $E_{*}$.
Thus, we consider the purification via the optimal environment $E_{*}$
and the postselection channel $K_{\omega}\otimes\mathbb{I}^{\text{E}*}$,
see Fig.~\ref{fig:purification}. Applying Eq.~(\ref{eq:meas-ineq-omg})
for the composite system consisting of the system, the optimal environment,
and the ancilla, we obtain 
\begin{equation}
I_{\omega}^{Q}[\sigma^{\text{SE}_{*}\text{A}}(x)]\leq I_{\omega}^{Q}[\ket{\Psi^{\text{SE}_{*}}(x)}],
\end{equation}
where 
\begin{equation}
I_{\omega}^{Q}[\sigma^{SE_{*}\text{A}}(x)]=I^{\mathrm{cl}}[p(\omega\big|x)]+p(\omega\big|x)I^{Q}[\ket{\Psi^{\text{SE}_{*}}(x|\omega)}],
\end{equation}
\begin{equation}
p(\omega\big|x)=\text{Tr}[\rho^{\text{SE}_{*}}(x)E_{\omega}\otimes\mathbb{I}^{\text{E}*}]=\text{Tr}[\rho(x)E_{\omega}],
\end{equation}
\begin{equation}
\ket{\psi^{\text{SE}_{*}}(x|\omega)}=M_{\omega}\otimes\mathbb{I}^{\text{E}*}\ket{\Psi^{\text{SE}_{*}}(x)}/\sqrt{p(\omega\big|x)},
\end{equation}
 and 
\begin{equation}
I_{\omega}^{Q}[\ket{\Psi^{\text{SE}_{*}}(x)}]=4\braket{\mathcal{D}_{x}\Psi^{\text{SE}_{*}}(x)\big|E_{\omega}\otimes\mathbb{I}^{\text{E}*}\big|\mathcal{D}_{x}\Psi^{\text{SE}_{*}}(x)}.
\end{equation}
Using the Uhlmann's theorem once again, we know 
\begin{equation}
I^{Q}[\ket{\Psi^{\text{SE}_{*}}(x|\omega)}]\ge I^{Q}[\sigma(x|\omega)],
\end{equation}
where 
\begin{equation}
\sigma(x|\omega)=\text{Tr}_{\text{E}_{*}}\left(\ket{\Psi^{\text{SE}_{*}}(x|\omega)}\bra{\Psi^{\text{SE}_{*}}(x|\omega)}\right).
\end{equation}
This leads to a similar version of Eq.~(\ref{eq:meas-ineq-omg}),
i.e.,
\begin{equation}
I_{\omega}^{Q}[\sigma^{\text{S}\text{A}}(x)]\leq I_{\omega}^{Q}[\ket{\Psi^{\text{SE}_{*}}(x)}]\label{eq:meas-ineq-omg-mixed}
\end{equation}
Summing over both sides over $\omega$, we know 
\begin{equation}
I^{Q}[\sigma^{\text{SA}}(x)]\leq I^{Q}[\ket{\Psi^{\text{SE}_{*}}(x)}]=I^{Q}[\rho(x)]\label{eq:meas-ineq-mixed}
\end{equation}

From this derivation, one can clearly see the challenging in postselected
quantum metrology with mixed states: the saturation of Eq.~(\ref{eq:PS-ineq})
in this case requires the exact knowledge of the optimal environment
$E_{*}$, which in general is formidable to find~\citep{escher2012quantum}.
On the other hand, without introducing the optimal environment, one
may need to evaluate $I^{Q}[\rho(x)]$ and $I^{Q}[\sigma^{\mathrm{SA}}(x)]$
respectively then check when they coincide. However, both $\rho(x)$
and $\sigma(x|\omega)$, the computation of corresponding QFIs involves
the cumbersome calculations of the SLD operators. 

As a result, due to these challenges, there is no systematic theoretical
framework on lossless postselection on mixed states in the current
literature. Physically, it is not even known whether post-selection
on mixed states can be still made lossless or not.

\refstepcounter{suppnote}

\section*{Supplementary Note 2: Proof of Theorem~\ref{thm:QP-PS}}

\label{SI:QP-PS-Proof}
\begin{proof}
Since any unitary transformation that is independent of the estimation
parameter will not change the QFI, without loss of generality, we
can simply take $U_{\checkmark}(x_{*})=\mathbb{I}$ and $M_{\checkmark}(x_{*})=\Pi_{s}(x_{*})+\sqrt{\lambda}_{\checkmark}\Pi_{r}(x_{*})$
and assume the limit $x_{*}\to x$. The post-selected state then is
\begin{equation}
\sigma(x|\checkmark)=\frac{M_{\checkmark}(x_{*})\rho(x)M_{\checkmark}^{\dagger}(x_{*})}{\text{Tr}\left[M_{\checkmark}(x_{*})\rho(x)M_{\checkmark}^{\dagger}(x_{*})\right]},\label{eq:sigPS}
\end{equation}
Therefore
\begin{align}
\partial_{x}\sigma(x|\checkmark) & =\frac{M_{\checkmark}(x_{*})\partial_{x}\rho(x)M_{\checkmark}^{\dagger}(x_{*})}{\lambda_{\checkmark}}\nonumber \\
 & -\frac{M_{\checkmark}(x_{*})\rho(x)M_{\checkmark}^{\dagger}(x_{*})}{\lambda_{\checkmark}^{2}}\text{Tr}\left[M_{\checkmark}(x_{*})\partial_{x}\rho(x)M_{\checkmark}^{\dagger}(x_{*})\right].\label{eq:dsigPS-dx}
\end{align}
Our goal is to show that upon imposing the ansatz given by Eq.~(\ref{eq:DO-derivative})
leads to the quasi-pure condition. 

With Eq.~(\ref{eq:drho-dx}), it is then straightforward to see $\Pi_{k}(x)\partial_{x}\rho(x)\Pi_{k}(x)=0$
holds for all density operators $\rho(x)$. Furthermore, it can be
readily checked that 
\begin{equation}
\text{Tr}[E_{\checkmark}(x)\partial_{x}\rho(x)]=\lambda_{\checkmark}\partial_{x}\left[\sum_{n=1}^{d_{r}}q_{n}(x)\right]=0,
\end{equation}
where $E_{\checkmark}(x)$ is defined in Eq.~(\ref{eq:QP-E-kernel})
and we have used the global normalization constraint~(\ref{eq:global-constr}).
Given above facts, Eq.~(\ref{eq:dsigPS-dx}) becomes 
\begin{align}
\lim_{x_{*}\to x}\partial_{x}\sigma(x|\checkmark) & =\frac{1}{\sqrt{\lambda_{\checkmark}}}[\Pi_{k}(x)\partial_{x}\rho(x)\Pi_{r}(x)+\text{h.c.}]\nonumber \\
 & +\frac{1}{\lambda_{\checkmark}}\Pi_{k}(x)\partial_{x}\rho(x)\Pi_{k}(x)+\Pi_{r}(x)\partial_{x}\rho(x)\Pi_{r}(x).
\end{align}
Now imposing Eq.~(\ref{eq:DO-derivative}), we know 
\begin{equation}
\partial_{x}\rho(x)=\left[\Pi_{k}(x)\partial_{x}\rho(x)\Pi_{r}(x)+\text{h.c.}\right]+\sqrt{\lambda_{\checkmark}}\Pi_{r}(x)\partial_{x}\rho(x)\Pi_{r}(x)
\end{equation}
On the other hand, we know from resolution of identity
\begin{equation}
\partial_{x}\rho(x)=\left[\Pi_{k}(x)\partial_{x}\rho(x)\Pi_{r}(x)+\text{h.c.}\right]+\Pi_{r}(x)\partial_{x}\rho(x)\Pi_{r}(x),
\end{equation}
Therefore the constraint~(\ref{eq:DO-derivative}) leads to the following
conditions: 
\begin{align}
\left(1-\sqrt{\lambda_{\checkmark}}\right)\Pi_{r}(x)\partial_{x}\rho(x)\Pi_{r}(x) & =0,
\end{align}
Since we exclude the trivial case where $\lambda_{\checkmark}=1$,
which correspond to applying identity postselection measurements,
we end up with the quasi-pure conditions.
\end{proof}
\refstepcounter{suppnote}

\section*{Supplementary Note 3: Details about the criteria of a quasi-pure
structure}

\label{SI:Criteria}

\subsection*{Proof of Observation~\ref{obs:GenCoDer}}
\begin{proof}
By definition, if $\ket{\psi_{k}(x)}$ and $\ket{\psi_{l}(x)}$ belongs
to the same degenerate subspace, then by definition $\ket{\mathscr{D}_{x}\psi_{n}(x)}$
is orthogonal to the degenerate subspace and therefore $\braket{\mathscr{D}_{x}\psi_{k}(x)|\psi_{l}(x)}=0$
holds. If $\ket{\psi_{k}(x)}$ and $\ket{\psi_{l}(x)}$ belongs to
different degenerate subspaces, then $\braket{\mathscr{D}_{x}\psi_{k}(x)|\psi_{l}(x)}=\braket{\partial_{x}\psi_{k}(x)|\psi_{l}(x)}=0$
\end{proof}

\subsection*{Proof of Observation~\ref{obs:QP-Conv}}
\begin{proof}
since $\{\ket{\psi_{k}(x)}\}_{k=1}^{w}$ spans $\mathcal{H}_{r}$,
they must be a linear combination of the set of the orthonormal of
basis $\{\varphi_{n}(x)\}_{n=1}^{d_{r}}$, Eq.~(\ref{eq:state-structure-spectral})
clearly implies Eq.~(\ref{eq:state-structure-convex}). On the other
hand, upon using the Gram-Schmidt orthogonalization process, it is
straightforward to see that Eq.~(\ref{eq:state-structure-convex})
implies Eq.~(\ref{eq:state-structure-spectral}).
\end{proof}
\refstepcounter{suppnote}

\section*{Supplementary Note 4: Proofs of the properties of quasi-pure states}

\label{SI:Properties}

\subsection*{Proof of Observation~\ref{obs:QP-SLD}}

Using the definition of the SLD~(\ref{eq:SLD-def}), we find 
\begin{equation}
\braket{\varphi_{k}(x)|\partial_{x}\rho(x)|\varphi_{l}(x)}=[p_{k}(x)+p_{l}(x)]\braket{\varphi_{k}(x)|L(x)|\varphi_{l}(x)}=0.
\end{equation}
Since $p_{k}(x)$ and $p_{l}(x)$ are strictly positive, the quasi-pure
condition also implies 
\begin{equation}
\braket{\varphi_{k}(x)|L(x)|\varphi_{l}(x)}=0,\,\forall k,\,l=1,2,\cdots w.
\end{equation}
Therefore the full expression of the SLD~\citep{yang2019optimal}
simplifies to 
\begin{equation}
L(x)=2\sum_{n}\left(\mathbb{I}-\Pi_{r}\right)\ket{\partial_{x}\varphi_{n}(x)}\bra{\psi_{n}(x)}+\mathrm{h.c.}
\end{equation}
On the other hand, thanks to the quasi-pure condition~(\ref{eq:QP-cond2}),
we know that $\Pi_{r}(x)\ket{\partial_{x}\varphi_{n}(x)}=\Pi_{n}(x)\ket{\partial_{x}\varphi_{n}(x)}$
and therefore 
\begin{equation}
L(x)=2\sum_{n}\ket{\mathscr{D}_{x}\varphi_{n}(x)}\bra{\psi_{n}(x)}+\mathrm{h.c.}
\end{equation}

\subsection*{Proof of Observation~\ref{obs:QP-QFI}}
\begin{proof}
Upon inserting Eq.~(\ref{eq:GenCoDer}) into the expression of the
QFI $I^{Q}[\rho(x)]=\mathrm{Tr}[\rho(x)L^{2}(x)]$, one immediately
obtain Eq.~(\ref{eq:QP-QFI}). Next, for non-degenerate $\rho(x)$,
the generalized covariant derivative becomes the conventional ones.
Furthermore, we note that $I^{Q}[\ket{\varphi_{n}(x)}]=4\braket{\mathcal{D}_{x}\varphi_{n}(x)|\mathcal{D}_{x}\varphi_{n}(x)}$,
which proves Eq.~(\ref{eq:QP-non-degenerate}). Finally, it is readily
checked that with the spectral decomposition of the mixed quasi-pure
states, $I^{\mathrm{cl}}[q_{n}(x)]=0$ due condition (i) in Observation~(\ref{obs:QP-spec}).
Then it immediately follows from Eq.~(\ref{eq:QP-non-degenerate})
that the inequality~(\ref{eq:Gen-Conv}) is saturated. 
\end{proof}

\subsection*{Proof of Observation~~\ref{obs:QP-PCC}}
\begin{proof}
Upon noting that 
\begin{equation}
L(x_{i})L(x_{j})=4\sum_{n}\ket{\mathscr{D}_{x_{i}}\psi_{n}(\bm{x})}\bra{\mathscr{D}_{x_{j}}\psi_{n}(\bm{x})}+\sum_{n,\,m}\ket{\psi_{n}(\bm{x})}\braket{\mathscr{D}_{x_{i}}\psi_{n}(\bm{x})|\mathscr{D}_{x_{j}}\psi_{m}(\bm{x})}\bra{\psi_{m}(\bm{x})}
\end{equation}
we find 
\begin{equation}
\braket{\psi_{k}(\bm{x})|L(x_{i})L(x_{j})|\psi_{l}(\bm{x})}=\braket{\mathscr{D}_{x_{i}}\psi_{k}(\bm{x})|\mathscr{D}_{x_{j}}\psi_{l}(\bm{x})}
\end{equation}
which completes the proof. 
\end{proof}

\end{document}